# Theoretical and Experimental Evidences of Material Phase Causality. Dynamics-Statistical Interpretation of the Quantum Mechanics.

**I. G. Koprinkov**

Department of Applied Physics, Technical University of Sofia, 1756 Sofia, Bulgaria

e-mail: igk@tu-sofia.bg

**Abstract.** The internal phase dynamics of a quantum system is revealed in details. Theoretical and experimental evidences of existence of a causal relation of the phase of the wave function with the dynamics of the quantum system are presented sistematically for the first time. A new, dynamics-statistical interpretation of the quantum mechanics is introduced. A particle-wave duality picture incorporated in the wave function through its phase and amplitude is considered.

## 1. Introduction

The physical state in the quantum mechanics is described not by a definite number of dynamical variables, *e.g.*, coordinates, linear momenta, etc., as in the classical mechanics, but by an abstract quantity – the state vector, or the wave function in a given representation, which obeys the Schrödinger equation [1]. Although the wave function describes the state of localized physical objects, *i.e.*, particle or system of particles, and the Hamiltonian of the system is written usually tacitly assuming point-like particles, the wave function itself, and, thus, the quantum state, is not localized, in general, but is distributed in space as for the continuous objects, *i.e.*, the waves. The wave-like nature of such point-like particle dynamics is further enforced by the validity of the quantum superposition principle and the quantum mechanical interference due to the linearity of the quantum evolution equation. Two ways of evolution of the wave function are known. If no measurement takes place, the wave function is subject to continuous, local, unitary (assuming Hermitian Hamiltonian), causal evolution ruled by the Schrödinger equation. The quantum state may evolve into a linear/coherent superposition of eigenstates of given dynamical observable and may reveal various wave-like features. During the measurement, the wave function undergoes a sudden, discontinuous, non-local, probabilistic collapse, the von Neumann's *process 1* [2], onto some of the eigenstates of the observable being measured selected by the measuring apparatus (or the environment), a process called decoherence. Although the collapse "shrinks" the state, it still remains distributed in space in view of the non-locality of the eigenstates of the quantum system. Thus, the particle-wave duality is built in the very nature of the quantum mechanics. The particle-wave duality and the nonlocality of the quantum phenomena are clearly demonstrated by the Feynman's "which-way" interference type experiments [3], Einstein-Podolsky-Rosen (EPR) type correlations [4], etc. The inherent duality of the quantum mechanical description predetermines the consequent problems with the interpretation of some results, sometimes considered as paradoxes. The description of the state of a localized physical object by a substantially non-local quantity, the wave function, does not have an analog in the classical physics. That is why, the wave function does not possess an obvious physical meaning that can be determined, *e.g.*, applying the correspondence principle between the classical and the quantum physics, and its relation to the physical reality is subject to additional considerations. Thus, the question of the relation between the wave function and the physical reality, and, on that ground – its physical meaning, becomes a fundamental problem in the quantum mechanics.

Since Einstein, in his famous debate with Bohr, has not succeed to present convincing arguments that "*God does not play dice*", the so called standard, or Copenhagen interpretation of the quantum mechanics has been widely accepted. It defeats the early proposed ontological point of view, according to which the wave function represents an objectively existing wave, *e.g.*, the de Broglie's pilot-wave associated with the particle. At the introduction of the wave function, the ontological meaning has been supported by Schrödinger. The Copenhagen interpretation encompasses several principles and conventions: the convention of the epistemological meaning of the wave function, the Heisenberg uncertainty principle, Bohr's complementary principle, etc. Within the Copenhagen interpretation, the wave function is only a tool in the determination of the probability to find a quantum system in a given state. In some aspect, the wave function is considered as the observer's knowledge (epistemology) about the state of the physical object.

Beside of the Copenhagen interpretation, a number of other interpretations have been proposed. Some of them involve substantial additional concepts and should be considered as new formulation rather than interpretation of the standard quantum mechanics. Among the most famous non-Copenhagen interpretations/formulations of the quantum mechanics are: de Broglie-Bohm ontological interpretation [5], Feynman's path integral formulation [6], consistent histories interpretation [7], many-worlds interpretation [8], environment-induced superselection [9], etc. In the ontological de Broglie-Bohm interpretation, one assumes that not only the particle (the primary object of the quantum mechanics) but also the wave-function (*i.e.*, the state of the particle, as in the conventional concept) represents a real physical object – a physical field. The evolution of that field is described by the Schrödinger equation, while the evolution of the particle position (considered as a *hidden variable* within the Bohm's theory) and its trajectrory is ruled by the physical action. An alternative approach to the quantum trajectories is developed based on the Feynman's path integral. In the many-worlds interpretation, the collapse of the wave function at the process of measurement, as in the Copenhagen interpretation, is replaced by decoherence of the quantum state at its interaction with the environment. At the same time, an assumption of existence of many parallel worlds is introduced, so that every possible outcome of the experiment is realized in some of these worlds. The consistent histories interpretation is based on the concept of finite ordered sequence of events taken at successive times, called history. A criterion of consistency allows selecting histories that have physical meaning. A complementary view to the quantum decoherence is the environment-induced superselection. It originates from the continuous decoherence of the states of the quantum system caused by the environment. This selects a small set of substates from the Hilbert space, which remain stable under further interaction with the environment. The above interpretations introduce some important concepts in quantum mechanics but their application still remains limited.

The Copenhagen interpretation confers a probabilistic [10] physical meaning to the amplitude of the wave function, only. In contrast to the amplitude, the role of the phase of the wave function is strongly neglected. Thus, for example, the time dependent phase, which appears in a factor of unit modulus, $\exp(-i\lambda t / \hbar)$, is considered, in principle, as unobservable [2], or, else, any true constant phase factor $\exp(i\alpha)$ is considered as irrelevant to any physical results [11]. Such an irrelevance of the phase of the wave function to the physical reality is subject to convention and, to the best of our knowledge, it has never been investigated and proved in an especially designed experiment prior (and after) the acceptance of the Copenhagen interpretation. In fact, the Copenhagen interpretation is formulated based on limited experimental evidences and knowledge about the quantum phenomena, obtained by rather simple apparatuses. A number of *gedanken* experiments [3, 4] have been proposed and exploited to analyze the behavior of the quantum objects that have not been directly accessible to experimental verification at that time. Nowadays, very refined laboratory equipment and experimental methods exist that offer an unprecedented precision in the performance of the experiments. This allows testing experimentally some of the foundations of the quantum mechanics as well as to realize some of the *gedanken* experiments [12-15]. Analyzing some experiments, one may clearly distinguish observable appearances of the material phase (MP) in the respective physical phenomena [16-19]. Some theoretical considerations and critical inspection of experiments with material wave-packets reveal that the relation of the wave function to the physical reality is not restricted to the amplitude only, but its phase is also strongly involved [20-22]. Here we present for the first time systematic theoretical and experimental evidences showing that the MP is causally related with the dynamics of the quantum system that may lead to observable physical consequences. The internal dynamics of the phase of a quantum system is revealed in details for the first time. All considerations are done in the frame of the quantum mechanics while avoiding the restrictions of the Copenhagen interpretation. The conclusions found here are used as a base of a more consistent, to our opinion, interpretation of the quantum mechanics, dynamics-statistical interpretation, according to which, the dynamical and statistical properties simultaneously take place in the quantum mechanical description of the physical reality.

## 2. Preliminaries and definitions

In the beginning, we will specify some terminological questions and definitions.

*First*, considering the MP, we have in mind the dynamical phase. Observable consequences of the geometric Berry phase [23] due to a cyclic parametric evolution of the Hamiltonian, *e.g.*, the Aharonov-Bohm effect, are known but such evolution and, thus, the Berry phase are not considered here.

*Second*, the constant/initial and the time dependent parts of the MP are considered together at an equivalent ground because of a number of arguments [20]: *i)* they have equivalent appearance in the initial and the subsequent mathematical expressions, *ii)* they both behave as constants within the (Hilbert) space of the states, whose vectors are functions on the configuration space, only, *iii)* changing the phase even by a constant value has a dramatic effect on the interference phenomena with material wave packets and it is accessible to experimental observation [16-19], *iv)* the initial phase, if retained in the wave function, shows sensible physical behavior, see Sec. 3.1, *v)* the initial phase represents actually the time-dependent phase acquired by the quantum system during its preceding history. In fact, the constant and the time depended phases are terms introduced by us. The interference phenomena do not make difference between those two phases. They simply feel any change of the phase does not matter by a constant or a variable value.

*Third*, we may distinguish the following *characteristic moments* of given quantum state: moment of creation of the state $t^*$ (usually, it is not considered within the quantum mechanics), initial moment $t_0$, *i.e.*, the moment from which the dynamical behavior of the quantum system is studied (it is considered with regard to the initial conditions that, however, usually not concern the MP), and, of course, the current moment of time $t$. The moment of creation of the state is the moment from which the time of existence of the created state is to be counted. The exact determination of that moment is subject to suitable definition and convention because a transient time exists between the preceding and the subsequently created states of the quantum system. The preceding history and the time of creation of given state are not a matter of concern in the usual treatment of the quantum mechanical problems, partially because, until recently, such questions were beyond the reach of the experimental methods. That is why, the moment of creation and the initial moment do not generally coincide. The present achievements in the ultrafast laser technology allow tracing the internal dynamics of the quantum systems. These include the nuclear motion inside the molecules in the femtosecond time scale [17-19], and the electron motion inside the atoms, in the pico/femtosecond time scale for some highly excited Rydberg atomic states [16]. Since recently, tracing the attosecond time dynamics of intraatomic electrons seems feasible [24-26]. Hence, the ability to trace the atomic and molecular transients gives sense to the question of the moment of creation of given quantum state, and, on that ground, the initial phase of the state.

*Forth*, in relation with the characteristic moments, we may distinguish the following parts of the dynamical phase. *Phase of creation* $\varphi_c(t^*)$ - the phase at the moment of creation $t^*$ of given quantum state. This phase is fixed to the same extend to what extend the time of creation can be defined within a given convention. *Initial dynamical phase* $\varphi_i(t^*, t_0)$ - the phase acquired during the preceding history of given quantum state, *i.e.*, from the moment of creation, $t^*$, to the initial moment, $t_0$. This phase is a constant. The initial phase includes the phase of creation as its own initial phase. *Evolution dynamical phase*, $\varphi_e(t_0, t)$, is the phase acquired by the quantum system during its time evolution considered explicitly within given physical problem, *i.e.*, from the initial moment of time $t_0$ to some arbitrary future moment of time $t$. This phase is, of course, a function of time. *Total dynamical phase*, $\varphi_t(t^*, t_0, t)$, is a sum of the initial phase and the evolution phase, *i.e.*, $\varphi_t(t^*, t_0, t) = \varphi_i(t^*, t_0) + \varphi_e(t_0, t)$. The total dynamical phase can be made well defined in an absolute sense and can be naturally considered as *absolute dynamical phase*, $\varphi_a(t^*, t_0, t) \equiv \varphi_t(t^*, t_0, t)$. This results from the fact that it includes the entire history of the quantum state, beginning from the moment of its creation. In view of such definition, the absolute phase keeps its absolute meaning even if, say, due to the methods of its practical determination, is referred to some other (reference) phase. Due to same reasons as for the characteristic moments, the practical determination of the MP is subject to suitable convention and experimental methods. It is also worth mentioned that, until recently, similar situation took place with the initial and, thus, with the total optical phase. It has been shown [27], however, that the initial (constant) optical phase (the accepted term in the respective field is absolute carrier-envelope phase) has well observable physical consequences. This allows exact determination and control of that optical phase with subsequent strong impact on the control of some high-field ultrafast phenomena. In the optical case, the initial/constant carrier-envelope phase is referred to the envelope/amplitude maximum. Similar approach can be used for the case of matter waves, i.e., the initial phase can be referred to some characteristic point of the amplitude of the wave function. Due to the direct relation between the dynamical phase and the physical action, $\varphi = -\hbar^{-1} S$, same definitions hold for the respective parts of the physical action $S$.

*Fifth,* we recognize the importance not only of the relative phase (*i.e.*, the phase difference between the absolute phases of the quantum states) but the absolute phase, itself. Knowledge the absolute value of the MP refers the problem to the absolute internal clocking of the state of the quantum system. Since this problem seems intractable using ordinary laboratory equipment (while not meaningless for the advanced experimental methods, as was said above), it is usually considered the relative phase, *i.e.,* the phase difference, if, by exception, the MP is discussed at all [2, 28, 29]. The importance of the initial constant phase and the evolution time dependent phase reflect, in fact, the importance of the initial conditions and the dynamical evolution, respectively, for the physical phenomena. The absolute (total) dynamical phase incorporates both kinds of factors. That is why, the absolute value of the dynamical phase is of primary importance in our considerations, while switching to the relative phase is trivial.

The initial and, thus, the absolute phase may look irrelevant to some particle-like phenomena but they have crucial importance for the wave-like interference phenomena. To recognize the role of the absolute dynamical phase is a matter of primary importance. The ways of extraction of physical information from the phase and its practical determination are, while also fundamental, but consequent problems. A way of extraction of the physical information from the MP/action is demonstrated in the Bohmian mechanics [5].

The MP causality will be considered as a particular case of a more general understanding of causality that will be introduced here. To our understanding, the causality is related to the existence of *fundamental physical relations* between *the objects of the physical reality*: particles, fields, and hierarchic constructs from them. The relations between the objects of the physical reality impose definite relations between the *elements of the physical reality* and the corresponding to them *physical quantities*. According to Einstein's point of view, "If, without in any way disturbing a system, we can predict with certainty (*i.e.*, with probability equal to unity) the value of the physical quantity, then there exists an element of physical reality corresponding to this physical quantity" [4]. Undoubtedly, this is the ideal for most of the physicists but its applicability is disputable from what is known from the present state of the theory and the experiment. In particular, it invokes the problem of the type of causality and determinism in the quantum mechanics [30]. In the present work, the elements of the physical reality will be recognized based on the fundamental relations that exist between the objects of the physical reality. Such relations are undisputable and exist irrespectively on the way of description of the physical reality. The contemporary physics reveals the existence of two different kinds of relations between the physical objects. The first one is based on the usual fundamental interactions. The second one involves EPR type correlations, and, according to Einstein, can be considered as a "spooky interaction at a distance". It may cause relation between the physical objects that are separated by a space-like interval, in sharp contrast to the requirement of local realism [4]. The EPR type correlations appear proved experimentally [12, 13] while their physical mechanism remains unclear. According to Einstein's understanding of a complete theory [4], "every element of the physical reality must have a counterpart in the physical theory". This again reflects the ideal of the orthodoxly thinking physicists, but it is hardly to believe that all elements of the physical reality sometime will be known. In fact, we do not know at present all elements of the physical reality. That is why, the accent will be put here on the physical quantities rather than on the related, sometimes abstract, elements of the physical reality. The physical quantities and their relationships are so important that they, in fact, constitute the essence of our knowledge about the physical objects and the physical phenomena. The meaningful physical behavior of given physical quantity is an indication of its relation with the physical reality. Sometimes, namely a particular behavior of given quantity (*e.g.*, conservation of momentum or energy) is a base of discovery of new physical objects. The following *definition of physical reality - physical quantity* relationship will be used here [20]: *If we find that a change (direct or indirect) of given quantity affects at least one other quantity, subject of direct or indirect observation/measurement, then the quantity under consideration is causally related with the physical reality. The quantity that is causally related with the physical reality will be called physical quantity, and* (if necessary, the latter can be extended in accordance to [4]) "...*there exists an element of physical reality corresponding to this physical quantity*". *On the same ground, the affected quantity is, of course, also a physical quantity.* Object/element of the physical reality (if one exists at all) that is not related to any other object/element of the physical reality is, in principle, unobservable and it seems meaningless to speak about it. Such kind of causality can be relatively easy and unambiguously established and this, to our opinion, is the most important and general problem that should be solved about given quantity. To such a quantity, after (if necessary) some additional studies and considerations, can be ascribed a definite *physical meaning.* The ultimate (at given stage of the science) qualitative and quantitative relationships between the physical quantities and their correspondence to the respective physical objects can be completely clarified within given physical theory. It must be underlined that both understandings, the present one and that one of Ref. [4], do not contradict each other in the sense that the requirement of relationship between the physical objects and the related physical quantities does not exclude the possibility that such a relationship allows predicting with certainty the values of the physical quantities. The opposite, however, does not hold, *i.e.*, a relationship may exist even if the behavior of the physical quantities cannot be predicted with certainty. Consequently, the present approach can be applied if any of the absolute (Laplacean) or the partial (quantum mechanical) [30] determinisms takes place in the physical reality.

The problem with the MP causality will be considered based namely on the above understanding of causality: *if a change of the MP causes observable physical consequences, e.g., change of other quantity that is subject of direct or indirect observation/measurement, then we may consider that the MP is causally related with the physical reality.* That is why, looking for a sensible physical behavior of the MP and its causal relationship with other observable physical quantities will be the main goal in the forthcoming considerations.

## 3. Evidences of material phase causality

The basic arguments in support of the MP causality will be classified in the following groups: special theoretical arguments, general theoretical arguments, and experimental evidences.

### 3. 1. Special theoretical arguments.

In this section we present the results of an extended study on the dynamics of the internal state of a quantum system, paying special attention to the MP [20]. To reveal the MP dynamics, the quantum system must be involved in a definite physical process. Isolated quantum system in stationary bare states is not suitable for that purpose. That is why, the quantum system will be a subject to interaction with an electromagnetic field and the environment. Nonadiabatic factors from the field (field time derivatives) and the environment (dumping) naturally appear in the rigorous solution [21] of the field-matter interaction problem.

The problem of the MP causality will be treated within an analytic solution $\left|\Psi(\vec{r},t)\right\rangle$ of the time dependent Schrödinger equation

$$\hat{H}\left|\Psi(\vec{r},t)\right\rangle = i\hbar\partial_t\left|\Psi(\vec{r},t)\right\rangle \qquad , \qquad (1)$$

applying a generalized adiabatic condition [20]. The Hamiltonian of the (two-level) quantum system under consideration, Fig.1, (in standard notations) is

$$\hat{H} = \sum_{j=1}^{2}\hbar\omega_j\left|j\right\rangle\left\langle j\right| - \mu E\left(\left|1\right\rangle\left\langle 2\right| + h.c.\right) - i\hbar\sum_{j=1}^{2}\frac{\gamma_j}{2}\left|j\right\rangle\left\langle j\right| \qquad . \qquad (2)$$

The original bare states $\left|1\right\rangle$ and $\left|2\right\rangle$ of the quantum system are coupled by a nonzero electric dipole moment. If $\left|1\right\rangle \equiv \left|g\right\rangle$ and $\left|2\right\rangle \equiv \left|e\right\rangle$ are ground and excited state, respectively, the dumping rate $\gamma_g$ of $\left|g\right\rangle$ can be taken zero. The states will be called "ground" $\left|1\right\rangle \equiv \left|g\right\rangle$ and "excited" $\left|2\right\rangle \equiv \left|e\right\rangle$ although the first one does not necessarily need to be a ground state in reality.

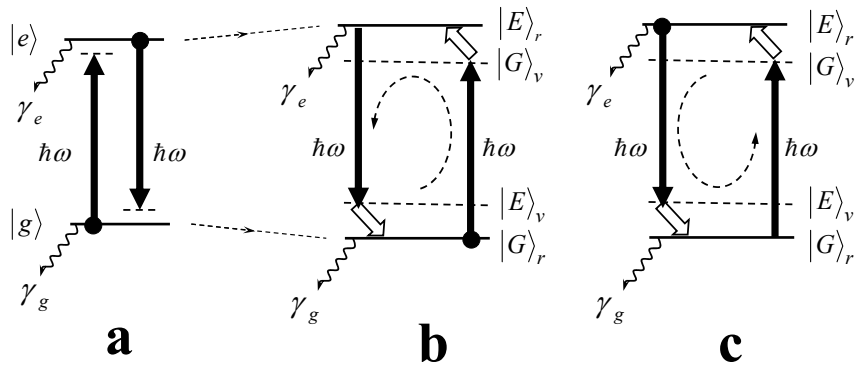

Fig.1. Bare (zero external field) (a) and dressed (non-zero external field) states (b) and (c) of the quantum system and the respective transitions at ground (b) and excited (c) state initial conditions. The bold arrows show the optical pumping, the hollow arrows show transitions due to the nonadiabatic factors, and the wavy arrows show the dumping. The black spots show the initial state (initial condition) from which the process of excitation and population of the states begins.

The electromagnetic (optical) field is expressed in the form

$$E = (1/2)\, E_o(t)\,[\exp(i\Phi_F(t)) + c.\,c.] \qquad , \qquad (3)$$

where $E_o(t)$ and $\Phi_F(t) = \omega t + \varphi(t)$ are the amplitude/envelope and the total phase of the field, respectively, $\omega$ is the carrier frequency of the field, and $\varphi(t)$ is the envelope-carrier phase. The above representation of the electromagnetic field in terms of envelope and carrier wave is quite general and holds even for ultrashort pulses of envelope time duration as short as the carrier wave period [27]. It is in the order of few femtoseconds for the electromagnetic fields in the optical range. The original bare states of the quantum system, from which the dressed states arise, are taken in the form [20]

$$\begin{aligned}
|g(\vec{r},t)\rangle &= |g(\vec{r})\rangle \exp\left[-i\,\Phi_g(t)\right] \\
|e(\vec{r},t)\rangle &= |e(\vec{r})\rangle \exp\left[-i\,\Phi_e(t)\right] \qquad ,
\end{aligned} \qquad (4)$$

where $\Phi_g(t) = \varphi_g + \omega_g\,t$, $\Phi_e(t) = \varphi_e + \omega_e\,t$ are the total dynamical phases; $\omega_g$, $\omega_e$ are eigenfrequencies/energies and $\varphi_g$, $\varphi_e$ are the respective initial phases. The latter are considered because of the arguments given in Sec. 2.

The state vector $|\Psi(\vec{r},t)\rangle$ is found as a solution of the time dependent Schrödinger equation (1) of Hamiltonian (2) and can be presented as

$$|\Psi(\vec{r},t)\rangle = \frac{1}{\sqrt{\widetilde{\Omega}'}}\left\{ \frac{\widetilde{\Lambda}'_{10}}{\sqrt{\widetilde{\Omega}'_0}} COS^{-1}\!\left(\frac{\theta}{2}\right)|G\rangle + \frac{\widetilde{\Lambda}'_{20}}{\sqrt{\widetilde{\Omega}'_0}} SIN^{-1}\!\left(\frac{\theta}{2}\right)|E\rangle \right\} \qquad , \qquad (5)$$

at *ground state initial conditions*, and

$$|\Psi(\vec{r},t)\rangle = \frac{1}{\sqrt{\widetilde{\Omega}'}}\left\{ \frac{\Omega_0}{2\sqrt{\widetilde{\Omega}'_0}} COS^{-1}\!\left(\frac{\theta}{2}\right)|G\rangle + \frac{\Omega_0}{2\sqrt{\widetilde{\Omega}'_0}} SIN^{-1}\!\left(\frac{\theta}{2}\right)|E\rangle \right\} \qquad , \qquad (6)$$

at *excited state initial conditions* [31].

New kind of states naturally arises from the initial bare states in the course of interaction of the quantum system with the electromagnetic field in presence of nonadiabatic factors. Such states of the combined "atom"-field system will be called *phase sensitive nonadiabatic dressed states* (PSNADSs) [32]. Each of the PSNADSs, ground $|G\rangle$ and excited $|E\rangle$, consists of real, $|G\rangle_r$ and $|E\rangle_r$, and virtual, $|G\rangle_v$ and $|E\rangle_v$, components

$$|G\rangle = COS\!\left(\frac{\theta}{2}\right)|G\rangle_r + SIN\!\left(\frac{\theta}{2}\right)|G\rangle_v \qquad (7)$$

$$|E\rangle = COS\!\left(\frac{\theta}{2}\right)|E\rangle_r - SIN\!\left(\frac{\theta}{2}\right)|E\rangle_v \qquad . \qquad (8)$$

The quantities $COS(\theta/2) = \sqrt{\widetilde{\Lambda}'_1/\widetilde{\Omega}'}$ and $SIN(\theta/2) = \sqrt{-\widetilde{\Lambda}'_2/\widetilde{\Omega}'}$ are complex functions that correspond to the usual $\cos(\theta/2)$ and $\sin(\theta/2)$ in the case of adiabatic dressed states [33], $\widetilde{\Omega}'$ is generalized (nonadiabatic) Rabi frequency. The subscript "0" in the respective quantities means their initial value at the initial moment $t = 0$. The real and the virtual components of the PSNADSs can be expressed in the following form:

$$\begin{aligned}
|G\rangle_r &= |g\rangle \exp(-i\Phi_{G,r}) \\
|G\rangle_v &= |e\rangle \exp(-i\Phi_{G,v}) \\
|E\rangle_r &= |e\rangle \exp(-i\Phi_{E,r}) \\
|E\rangle_v &= |g\rangle \exp(-i\Phi_{E,v})
\end{aligned} \qquad , \qquad (9)$$

where $\Phi_{G,r}$, $\Phi_{G,v}$, $\Phi_{E,r}$, and $\Phi_{E,v}$ are the total phases of the respective real and virtual components of the dressed states.

At *ground state initial conditions*, the total phases of the dressed state components are:

$$\Phi_{G,r} = \varphi_g + \int_0^t \widetilde{\omega}'_G dt'$$

$$\Phi_{G,v} = \Phi_{G,r} + \Phi_F = (\varphi_g + \varphi(t)) + \int_0^t (\widetilde{\omega}'_G + \omega) dt'$$

$$\Phi_{E,r} = \Phi_{G,r} + \Phi_F + \Phi_{NAD} = (\varphi_g + \varphi(t)) + \int_0^t \widetilde{\omega}'_E dt'$$

$$\Phi_{E,v} = \Phi_{E,r} - \Phi_F = \varphi_g + \int_0^t (\widetilde{\omega}'_E - \omega) dt'$$

(10)

At *excited state initial conditions*, the total phases of the dressed state components are:

$$\Phi_{E,r} = \varphi_e + \varphi(t) - \varphi_0 + \int_0^t \widetilde{\omega}'_E dt'$$

$$\Phi_{E,v} = \Phi_{E,r} - \Phi_F = \varphi_e - \varphi_0 + \int_0^t (\widetilde{\omega}'_E - \omega) dt'$$

$$\Phi_{G,r} = \Phi_{E,r} - \Phi_F - \Phi_{NAD} = \varphi_e - \varphi_0 + \int_0^t \widetilde{\omega}'_G dt'$$

$$\Phi_{G,v} = \Phi_{G,r} + \Phi_F = \varphi_e + \varphi(t) - \varphi_0 + \int_0^t (\widetilde{\omega}'_G + \omega) dt'$$

(11)

Here, $\widetilde{\omega}'_G$ and $\widetilde{\omega}'_E$ are nonadiabatic frequencies/energies of the ground and the excited dressed states, respectively, $\Delta\widetilde{\omega}'_{NAD} = \widetilde{\omega}'_E - \widetilde{\omega}'_G - \omega$ is nonadiabatic frequency detuning, and $\Phi_{NAD} = \int_0^t \Delta\widetilde{\omega}'_{NAD} dt' = \int_0^t (\widetilde{\omega}'_E - \widetilde{\omega}'_G - \omega) dt'$ is nonadiabatic phase. The frequencies of the ground $\widetilde{\omega}'_G$ and excited $\widetilde{\omega}'_E$ dressed states are Stark shifted and modified by the nonadiabatic contributions from the field ($\partial_t \varphi$, $\Omega^{-1} \partial_t \Omega$) and the dumping ($\gamma_g$, $\gamma_e$). The mathematical details of the approach can be found in [21].

The behavior of the dynamical phases can be revealed analyzing Eqs. (9)-(11). Starting from $|G\rangle_r$, *ground state initial condition* or $|g\rangle - - \rightarrow |G\rangle_r \rightarrow |G\rangle_v \Rightarrow |E\rangle_r \rightarrow |E\rangle_v$ sequence (the dashed arrow shows continuous evolution of the initial bare state toward the respective real dressed state component, the continuous arrow shows radiative transition within given dressed state, and the hollow arrow shows nonradiative nonadiabatic transition between the different dressed states), Fig.1(b), and using Eqs. (9) and (10) one may find the following development of the states and their phases along with the relevant physical processes. The bare ground state $|g(\vec{r},t)\rangle$ evolves toward a new ground state, the dressed ground state $|G\rangle$, while the eigenfrequency/energy $\omega_g$ of $|g\rangle$ evolves continuously into the eigenfrequency/energy $\widetilde{\omega}'_G$ of the real component $|G\rangle_r$ of the ground state $|G\rangle$. At the same time, the electromagnetic field "peaks up" the initial phase $\varphi_g$ of the bare ground state and, modifying the instantaneous energy of the state, forces the total phase $\Phi_g$ of $|g(\vec{r},t)\rangle$ to evolve, forming in this way the total phase $\Phi_{G,r}$ of $|G\rangle_r$. Once $|G\rangle_r$ is created, the process of creation/population of the other dressed state components and their phases can be traced out starting from $|G\rangle_r$. The next step is formation of $|G\rangle_v$ from $|G\rangle_r$. The phase $\Phi_{G,v}$ of the virtual component $|G\rangle_v$ results from the phase $\Phi_{G,r}$ of the real component $|G\rangle_r$ adding the field (optical) phase $\Phi_F$. At the same time, physically, the virtual component of the ground state results from the real component of the ground state by temporal association (and reemission) of one photon from (to) the field, stimulated by the field itself [21]. In the next step, formation of $|E\rangle_r$ from $|G\rangle_v$, the phase $\Phi_{E,r}$ of the real component $|E\rangle_r$ results from the phase $\Phi_{G,v}$ of the virtual component $|G\rangle_v$ adding the nonadiabatic phase $\Phi_{NAD}$. In this process, the quantum system acquires nonadiabatic contributions from the field and the dumping while it overcomes the nonadiabatic frequency detuning $\Delta\widetilde{\omega}'_{NAD}$. At the same time, in accordance with the adiabatic theorem of quantum mechanics [34], the physical transition between different states, in this case $|G\rangle_v$ and $|E\rangle_r$ components of the ground and excited dressed states, results namely from the nonadiabatic factors acting on the quantum system. This is accompanied by irreversible absorption of one photon from the field thus leading to continuous (within the lifetime of the state) population of the real excited state $|E\rangle_r$. Finally, the phase $\Phi_{E,v}$ of the virtual component $\left|E\right\rangle_v$ results from the phase $\Phi_{E,r}$ of the real component $\left|E\right\rangle_r$ subtracting the optical phase $\Phi_F$. At the same time, physically, the virtual component of the excited state results from the real component of the excited state by temporal emission (and reabsorption) of one photon to (from) the field, stimulated by the field itself [21]. The above analysis shows exact correspondence between the MP behavior and the respective physical processes. In particular, it naturally explains why the initial MP $\varphi_g$ of the ground state $\left|g\right\rangle$, from which the process of formation of the dressed states begins at ground state initial conditions, appears in all dressed state components, whereas the excited state initial MP $\varphi_e$ totally disappears. Similar consequences can be revealed at *excited state initial condition*, or $\left|e\right\rangle ---\rightarrow \left|E\right\rangle_r \rightarrow \left|E\right\rangle_v \Rightarrow \left|G\right\rangle_r \rightarrow \left|G\right\rangle_v$ sequence, Fig.1(c), using Eqs. (9) and (11). The bare excited state $\left|e(\vec{r},t)\right\rangle$ evolves toward a new excited state, the dressed excited state $\left|E\right\rangle$, while the eigenfrequency/energy $\omega_e$ of $\left|e\right\rangle$ evolves continuously into the eigenfrequency/energy $\tilde{\omega}_E^r$ of the real component $\left|E\right\rangle_r$ of $\left|E\right\rangle$. At the same time, the electromagnetic field "peaks up" the initial phase $\varphi_e$ of the bare excited state and, modifying the instantaneous energy of the state, forces the total phase $\Phi_e$ of $\left|e(\vec{r},t)\right\rangle$ to evolve, forming in this way the total phase $\Phi_{E,r}$ of $\left|E\right\rangle_r$. The formation/population of the dressed state components can be traced in a similar way as in the case of the ground state initial conditions, starting from $\left|E\right\rangle_r$ in this case. The next step is formation of $\left|E\right\rangle_v$ from $\left|E\right\rangle_r$. The total phase $\Phi_{E,v}$ of $\left|E\right\rangle_v$ results from the total phase $\Phi_{E,r}$ of $\left|E\right\rangle_r$ subtracting the optical phase $\Phi_F$. Physically, the virtual component $\left|E\right\rangle_v$ results from the real component $\left|E\right\rangle_r$ by a temporal release (and reabsorption) of one photon to (from) the field. The transition from the virtual state $\left|E\right\rangle_v$ to the real state $\left|G\right\rangle_r$ results from the action of nonadiabatic factors, acquiring the respective nonadiabatic phase $\Phi_{NAD}$. This is accompanied by irreversible emission of one photon to the field leading to continuous (within the lifetime of the state) population of the real ground state $\left|G\right\rangle_r$. Finally, the total phase $\Phi_{G,v}$ of $\left|G\right\rangle_v$ results from the total phase $\Phi_{G,r}$ of $\left|G\right\rangle_r$ adding the field phase $\Phi_F$. The latter corresponds to the physical formation of $\left|G\right\rangle_v$ from $\left|G\right\rangle_r$ as a result of temporal association (and reemission) of one photon from (to) the field. Again, the initial phase $\varphi_e$ of the state from which the process begins, in this case - the excited state $\left|e\right\rangle$, appears in all dressed states components, while the ground state initial phase $\varphi_g$ totally disappears, Eq.(11). It is important to note that such behavior of the initial phase does not result from a kind of trivial elimination of the respective phases due to the initial conditions but represents a general behavior of the MP. It is a consequence of the specified way of formation of the respective quantum states and their phases.

Although the MP behavior is the main subject of this work, the particular behavior of the optical phase is also important for the phase dynamics of the quantum system because it explicitly participates in the formation of the total phase of the PSNADS. The appearance of the optical phase at ground state initial conditions is rather simple and transparent and it promotes understanding of the MP dynamics. Whereas the energy of the original ground state $\left|g\right\rangle$ evolves continuously to that one of $\left|G\right\rangle_r$, the initial phase $\varphi_g$ of $\left|g\right\rangle$ transfers without change and becomes the initial phase of $\left|G\right\rangle_r$, Eq. (10). At excited state initial conditions, however, the total phase of the PSNADS is subject to more complicated and non-intuitive way of formation. If the phase $\varphi(t)$ of the optical pulse remains constant, $\varphi(t) = \varphi_0 = const$, the term $\varphi(t) - \varphi_0$ in Eq.(11) cancels and the phase behavior at excited state initial conditions coincides with that one at ground state initial conditions. In that case, the initial phase $\varphi_e$ of the original bare excited state $\left|e\right\rangle$ again transfers without change and becomes the initial phase $\varphi_e$ of $\left|E\right\rangle_r$. If the phase $\varphi(t)$ changes in time, however, a nonzero phase difference $\varphi(t) - \varphi_0 = \sum_{k=1}(k!)^{-1}\varphi^{(k)}(t=0)t^k$ appears in the total phase $\Phi_{E,r}$ of $\left|E\right\rangle_r$ in the case of excited state formation sequence, Eq. (11). With exception of envelope-carrier phase $\varphi_0 = \varphi(t=0)$, the MP of the initial dressed state component $\left|E\right\rangle_r$ "absorbs" (in the course of its formation from the bare state $\left|e\right\rangle$) all time dependent phase contributions from the forcing optical field coming from carrier frequency shift, $\varphi_1 = \partial_t^1 \varphi(t=0)$, linear chirp, $\varphi_2 = \partial_t^2 \varphi(t=0)$, quadratic chirp, $\varphi_3 = \partial_t^3 \varphi(t=0)$, etc. It can be considered as a kind of nonadiabatic penetration of the time dependent optical phase into the phases of the PSNADSs. In the light of the above considerations, we may also speculate that it is the phase/action that is actually added in response to the physical phenomena whereas the additive behavior of the frequency/energy (the time derivative of the phase/action) appears to be a consecutive result.

The dynamics of the quantum system within the PSNADSs shows that *the MP behaves as an additive dynamical quantity that causally follows the initial conditions (the initial, constant part of the MP) and the physical process (the evolution time dependent part of the MP) responsible for the formation/population of given quantum state*. Such a behavior of the MP has been called MP tracking [20]. The MP tracking takes place within both, the adiabatic [35] and the nonadiabatic [20] dressed states. This means that it is more fundamentally related to the dynamics of the quantum system than the particular way, adiabatic or nonadiabatic, of the field-matter interaction. The above analysis reveals not only the physical behavior of the MP but also that such a behavior cannot be well understood without taking into account the initial conditions and the relevant physical processes. To the best of our knowledge, such detail dynamics behavior of the material and the field phases is presented for the first time. Elements of the discussed above behavior of the MP can be distinguished in a number of experiments with material wave packets (for details, see section 3.3). The theoretically and experimentally observed manifestation of the causal dependence of the MP on the physical processes and the initial conditions is actually what it is usually considered as a *sensible physical behavior* of given quantity.

*3. 2. General theoretical arguments.*

In this section we present general theoretical arguments showing that the MP is causally related to the physical reality. In the polar representation, the wave function $\Psi = R(\vec{r},t)\exp[-i\Phi(\vec{r},t)] = R(\vec{r},t)\exp[iS(\vec{r},t)/\hbar]$ consists of amplitude $R(\vec{r},t)$ and phase $\Phi(\vec{r},t) = -\hbar^{-1}S(\vec{r},t)$, where $S$ is the physical action. In other words, up to a constant factor (determined by a fundamental physical constant $\hbar$), the MP coincides with the physical action. It can be shown that the amplitude $R$ and the action $S$, or the phase $\Phi$ of the wave function, are not independent but become mutually related. Such a relation can be revealed within the so called "hydrodynamic" representation of the quantum mechanics. The latter is known since the works of de Broglie [36] and Madelung [37]. Later on, it is extensively exploited by D. Bohm [5] in the ontological (causal) interpretation of the quantum mechanics based on hidden variables concept [4]. Substitution of the wave function in the Schrödinger equation leads to the following coupled differential equations for the action/phase and the amplitude

$$\partial S/\partial t + (\nabla S)^2/2m + V(\vec{r},t) - (\hbar^2/2m)(\Delta R/R) = 0 \qquad (12)$$

$$\partial (R)^2/\partial t + \nabla.(R^2 \nabla S/m) = 0 \qquad\qquad . \qquad (13)$$

The first equation (12) is known as quantum mechanical equation of Hamilton-Jacobi for the action/phase, $V(\vec{r},t)$ is the usual interaction potential, and $U(\vec{r},t) = -(\hbar^2/2m)(\Delta R/R)$ is the so called quantum potential. The second equation (13) is the continuity equation for the quantum probability density $\rho = R^2 = |\Psi|^2$ [38]. The quantum potential, according to the Bohm's interpretation of quantum mechanics, originates from a real interaction between the wave function $\Psi$, considered as "a objectively real field" [5], and the particle. This field exerts a force on the respective particle by means of the quantum potential and, together with the usual interaction potential, governs its propagation.

The hydrodynamic representation is used in the Bohmian mechanics to introduce a quantum concept of particle moving along a trajectory. Within this concept, an ensemble of Bohmian trajectories arises from the integration of the velocity $\partial_t \vec{r} = \vec{v} = m^{-1}\nabla S(\vec{r},t)$. In our work, the hydrodynamic representation is used not to exploit the concept of trajectory (although the latter is close to our understandings, see Appendix A) but to prove the MP causality. We neither consider the wave function as an objectively existing field, nor the quantum potential as a really existing potential that is capable to exert a force on a particle, see Sec.4.

The phase/action $\Phi/S$ and the amplitude $R$ obey coupled differential equations (12) and (13). Therefore, they are not independent but codetermine each other, i.e., $S/\Phi \leftrightarrow R$ relationship exists. Any change of the amplitude of the wave function influences the action/phase and vise versa. That is why, it is hardly to believe that the square of the amplitude $R^2 = |\psi|^2$ of the wave function is related to some element of the physical reality, while its action/phase $S/\Phi$ is irrelevant to the same reality. Taking into account $S/\Phi \leftrightarrow R$ relationship that follows from Eqs. (12), (13) and the widely accepted $R \leftrightarrow$ *physical reality* relationship (as, *e.g.*, in the Copenhagen interpretation, Bohm's ontological interpretation [5], etc.), existence of a complete

$$S/\Phi \leftrightarrow R \leftrightarrow physical\ reality \qquad\qquad (14)$$

relationship can be established. *Such $S/\Phi \leftrightarrow R \leftrightarrow physical\ reality$ relationship is fundamental and exists independently on the particular interpretation of the quantum mechanics*. The latter, however, can be considered as a basic reference point in the interpretation of the quantum mechanics. What is also important, it is namely the action/phase $S/\Phi$, but not the amplitude $R$, that is ruled by the real dynamical equation (12). The amplitude $R$ obeys the continuity equation (13), which plays only an auxiliary role with respect to the quantum dynamics, *i.e.*, to rule the conservation of the probability density. In view of the above considerations, the phase/action appears to be much more deeply related with the dynamics of the quantum system than the amplitude. Therefore, one may expect that much more fundamental physics is related namely with the MP. The logical relation $S/\Phi \leftrightarrow R \leftrightarrow physical\ reality$ is the most important, to our opinion, theoretical argument that proves the MP causality. It is completely general because it results solely from the structure of the Schrödinger equation and the wave function. The only the assumption involved here is the widely accepted (not only in the Copenhagen interpretation) relation between the amplitude of the wave function and the physical reality. Whereas the hydrodynamic representation of the quantum mechanics is well established, it has not been used till now (to the best of our knowledge) to reveal the existence of a fundamental relation between the phase of the wave-function and the physical reality.

It is usually considered that the constant action/phase does not matter for the physical phenomena [39] because the action participates only in derivatives in the classical and the quantum Hamilton-Jacobi equation (12), as well as in the continuity equation (13). Such a point of view is not supported here, especially if the interference phenomena are considered. The action/phase is an additive quantity and the action/phase acquired during the dynamical evolution is added to its initial (constant) part, which is determined by the initial conditions. The sensible physical behavior of the initial value of the phase/action has been already discussed in Sec. 2 and 3.1. What is more important, however, any constant shift of the phase/action may lead to observable effects at suitably designed (interference) experiments.

While the above argument is enough to prove the MP causality at a fundamental level, another general consideration worth also to be mentioned. It arises from the quantum superposition principle. While not explicitly included within the postulates of the quantum mechanics, *e.g.*, [29], it is one of the foundations of the quantum mechanics and constitutes the essence of the quantum mechanical interference. The interference between the quantum states, as well as between the quantum transitions, represents the very nature of the quantum phenomena. Having in mind the quantum superposition principle [11]

$$\Psi(\vec{r},t) = \sum_i C_i \Psi_i(\vec{r},t) \qquad\qquad , \qquad\qquad (15)$$

it is a trivial mathematical result that both, the amplitudes $|\Psi_i|$ and the phases $\Phi_i$ of the superimposed states $\Psi_i$ participate in the determination of the amplitude $|\Psi|$ and the phase $\Phi$ of the resultant state $\Psi$. Any change of the phases of the superimposed states may cause observable results related with the entire resultant state, *i.e.*, its amplitude and phase. This is another evidence of the mutual relation between the phase and the amplitude of the wave function.

*3. 3. Experimental evidences.*

The most important point in the present considerations is that the above theoretical arguments have experimental confirmations. In fact, a large number of experimental evidences from various areas of physics exist in support of the MP causality. Here, we will summarize the main conclusions that can be done from some of the most convincing, to our opinion, experimental studies, without pretending for completeness of this survey. The experiments under consideration have not been particularly designed to study the MP causality and critical inspection of the conditions and the experimental results are required in order to make correct conclusions.

The interference of the matter waves is a basic quantum mechanical phenomenon. A general outcome of the mater-wave experiments consists in the fact that, changing the MP, one may affect the interference picture with material wave-packets, *i.e.*, the population of given internal state of the quantum system, or the number density distribution of the particles in space, and all these are subject to experimental observation. Phase sensitivity of the observable quantities takes place with respect to, both, the field phase (the phase of the optical pulses that create the material wave-packets), which is a well recognized effect, and, what is particularly important here - the MP (the phase of the material wave-packet itself). The accent will be put on experiments that clearly demonstrate the MP dependence of the physical phenomena, while some important aspects of the optical phase dependence will also be considered.

The material wave packets can be classified into two well distinguishable pure types: *(i)* internal wave-packets (superposition of the bound states in a quantum system - atom, molecule) [16-19] and *(ii)* external, de Broglie wave-packets (external motion state of free or quasi-free particle - electron, neutron, atom, molecule, etc.) [14, 15]. To illustrate the discussion, the following rather general form of the MP will be considered:

$$\Phi(\vec{r}, t) = \hbar^{-1}\left(\int_{t_0}^{t} E \, dt' - \int_{\vec{r}_0}^{\vec{r}} \vec{p}.d\vec{r}'\right),$$ where $E$ is the energy and $\vec{p}$ is the linear momentum. The total MP can

be ruled either by control of its initial, constant part, or by control of its variable part. The initial phase depends on the physical conditions at the initial moment $t_0$ and the initial position $\vec{r}_0$. It can be manipulated, *e.g.*, by the phase of the electromagnetic field that creates the wave packet, see Eqs. (10) and (11) and the related discussion in Sec.3.1. The variable phase depends on the particular dynamics in which the quantum system becomes involved. It can be controlled changing the energy $E$ of the quantum system and/or the evolution time (from $t_0$ to $t$), or changing the linear momentum $\vec{p}$ and/or the evolution path (from $\vec{r}_0$ to $\vec{r}$) to the time/position of the interference with another, reference wave packet. For the case of internal wave packets, the energy of the quantum system can be changed by a selective excitation of a particular superposition of electronic states in an atom [16], or a particular superposition of rovibration states of given electronic state in a molecule [17-19]. Change of the linear momentum or the whole phase/action path integral can be realized by a selective excitation of given electron trajectory [40]. The creation of stable interference picture is subject to a kind of coherence between the matter waves, in a similar way as for the case of optical waves.

The phase sensitive experiments will be divided (for convenience in the forthcoming discussion) into the following types: *(i)* experiments with bound intraatomic/intramolecular wave-packets; *(ii)* experiments with quasi-free, tunnel electron wave-packets; and *(iii)* experiments with free wave-packets.

*(i) experiments with bound (intraatomic/intramolecular) wave packets*

The MP dependence of the physical observables can be distinguished in an analog of Young's double slit interferometer within an atom [16], or Michelson interferometer within a molecule [17-19]. In these experiments, the light beams in the usual optical interferometers are replaced by electron (within the atom) or nuclear (within the molecule) wave packets. The wave packets are created by a sequence of usually two (pump) laser pulses of controllable phase. A third (probe) laser pulse is used to probe the evolution of the superposition of the wave packets. To achieve well expressed MP effects, the wave packets and, thus, the pump pulses, must be substantially shorter than the electron orbiting time in the particular atomic state or the nuclear vibration time in the particular molecular state involved in the experiment. Once created, the wave packets are ruled solely by the intraatomic/intramolecular Hamiltonian in the time between pulses. After some propagation time, the wave-packets may overlap and interfere. The result of the interference can be detected, *e.g.*, by fluorescence interferogram from an excited state [17-19], or ion current due to ionization of the atom/molecule from the superposition state [16]. Beside of other factors, the local population of the interference state depends on the MP acquired by the wave packets during their evolution. The MP difference between the wave packets inside the atoms/molecules can be controlled by a number of ways: (*i*) changing the phases of the pump pulses, and, in this way, the initial phase of the created wave packets [16], (*ii*) changing the delay time between the pump pulses, and, in this way, the relative MP between the created wave packets [16-19], (*iii*) changing the carrier frequency of the pump pulses, which changes the mean Bohr frequency of the intraatomic wave packets, or the carrier vibration frequency of the intramolecular wave packets, and, thus, the time rate of the MP acquisition [17, 18], or (*iv*) any combination from above. The most important outcome from these experiments is that a change of the MP in some of the specified ways leads to observable physical effects: change of the fluorescence interferogram, the ion current, etc., [16-19]. In particular, even a constant phase shift of the phase-locked laser pulses that create the material wave packets leads to a dramatic change in the observed result [16–18]. In addition, the well defined behavior of the field-matter phenomena is a strong evidence that the material wave packet is able to "carry" in a deterministic way the optical phase, "absorbed" from the optical pulses at the process of creation of the wave packets, adding to it a definite value of the MP acquired during the evolution of the wave packets inside the quantum system (atom, molecule, etc.).

*(ii) experiments with quasi-free tunnel electron wave packets*

One of the most amazing advances in the ultrafast physics is the generation of single attosecond pulses by means of atomic level control of the emission of high-harmonics from tunnel electrons driven by high-intensity femtosecond laser pulses. The underlying physical processes of the high-harmonic generation are strongly sensitive to the particular trajectories of the tunnel electrons. Electron trajectory concept in the spirit of Feynman's path integral formulation of the quantum mechanics [6] becomes a very successful approach in the description of some high-field phenomena: high-harmonic generation, above-threshold ionization [40]. Within this approach, the action/phase acquired by the tunnel electrons along definite trajectories, quantum paths, plays a key role in the above phenomena. It leads to emissions that differ in strength, energy region [40], as well as in spatial and temporal coherence [41]. In this way, the contributions of the different trajectories can be experimentally distinguished, leading even to macroscopically visible results [41]. In the same way as for the PSNADS, Sec. 3.1, the created wave packet "absorbs" the optical phase and adding the acquired pure MP forms the total phase of the tunnel electron wave packet. Influence of the phase modulation (chirp) of the exciting pulse on the spectra of the generated high-harmonics [42] is another indication that the field and the matter (atomic dipole) phases add together in a predictable and controllable way. Changing the phase of the laser field and/or MP by means of selective excitation of definite electron trajectories affects the total, optical plus material, phase of the wave packet and it is accessible to experimental observation [40-42].

*(iii) experiments with free wave-packets*

Every experiment on interference between particles (matter waves) [13-15, 43-45] is, in fact, an example of phase sensitive phenomenon. Nevertheless, below we will point out only experiments which explicitly show that the interference picture is sensitive to *the change* of the MP.

Interference of the wave packets of free electrons is observed experimentally using extreme ultraviolet attosecond pulses, to ionize argon atoms, and a suitable infrared laser field, to induce a momentum transfer of the released atomic electrons [46]. Dependence of the interference picture on the phase of the electron wave packets, *i.e.*, the MP, can be distinguished in these experiments. The phase of the wave packets is manipulated by a momentum transfer from the infrared laser field, or changing the time delay between the attosecond and the infrared pulses. The change of the phase of the electronic wave packets leads to observable change in the interference picture [46].

The atomic/molecular interference is basic phenomenon in the atomic/molecular matter wave interferometry. Change of the phases of the atomic partial waves, and thus, the created (Ramsey) fringes, by means of (*i*) change of the phase of the laser fields that are used to split and recombine the atomic beam in a Ramsey interferometer, (*ii*) ac-Stark shift, or (*iii*) rotation of the atomic interferometer (Sagnac effect) is well known observable result in the matter wave interferometry [45, 47]. Thus, the matter-wave interferometry is another source of evidences of the dependence of the physical phenomena on the MP.

The above theoretical (Sec. 3.1, 3.2) and experimental (Sec. 3.3) evidences conclusively establish the existence of a fundamental relation of the MP with the physical reality and we may summarize that *the MP is causally related with the dynamics of the quantum system*. The latter, together with the fact that a Hermitian operator can be associated to the phase [48, 49], allows considering the MP as a physical observable on the same formal basis as any other physical observable. Hence, although the entire wave function $\Psi$ does not have real physical meaning, each of its elements, amplitude $R$ and phase $\Phi$ ($s$), has observable appearances in the physical phenomena. The MP causality brings the matter waves closer to the electromagnetic waves (Appendix B), for which the physical meaning of the phase is undisputable. In the light of above arguments, it seems apparent that the Copenhagen interpretation unnecessarily restricts the physical meaning of the wave function to its amplitude, only. Investigations on the MP behavior may lead to much deeper understanding of the intimate nature of the quantum phenomena (Appendix C).

## 4. Dynamics-statistical interpretation of the quantum mechanics

In spite of the well expressed relation between the physical reality and the MP, it yet remains irrelevant to the interpretation of the quantum mechanics. Some basic features of the MP announced here are not consistent with the Copenhagen interpretation, or some of the most significant non-Copenhagen interpretations [5-9]. This invokes considering of another interpretation. It will be based solely on the basic principles of the quantum mechanics, without involving additional hypotheses. Also, we hope that it is not influenced by the observer's prejudice about the physical reality.

Let we first begin with the ontological point of view according to which the wave function is an objectively existing field $\Psi$ that is capable to exert a force on the related particle by means of the quantum potential [5]. The quantum potential depends on the instantaneous positions $\vec{r}_1,...,\vec{r}_n$ of all particles in the system at given moment of time $t$, $U(\vec{r}_1,...,\vec{r}_n,t) = -\left(\hbar^2/2m\right)\Delta R(\vec{r}_1,...,\vec{r}_n,t)/R(\vec{r}_1,...,\vec{r}_n,t)$, and it is considered as a non-local connection between the individual physical objects of the system. The extrapolation of this approach to the universe leads to the Bohm's idea of existence of an implicate order and an undivided wholeness of the entire universe through its quantum potential. Such a physical picture, however, should be introduced in the theory since the very beginning with the subsequent dramatic reconstruction of the foundations of the physics. In that case, instead of motion and interaction between individual particles and fields (considered as relatively independent entities), the physical phenomena should also reflect the additional contribution of all material objects through the quantum potential of the system. If this is so, the ultimate solution of any particular problem would require the knowledge of the state of the entire universe, more particularly, the instantaneous positions of all particles in it. While it could be a correct approach, there are not enough evidences that it is namely the quantum potential that may play the role of such an instantaneous unification factor. A useful reference point when considering the problem with the interpretation of the quantum mechanics, in general, and the quantum potential, in particular, is to go back to the origins. The physical processes involved in given physical problem are specified in advance and it is, in fact, a precondition to construct the Hamiltonian in the Schrödinger equation and to solve the problem. Any material system consists of particles, each of which moves (finitely or infinitely) in space and, consequently, possesses kinetic energy described by the kinetic energy operator $\hat{T} = \hat{p}^2/2m$, and interacts with the other particles or the environment by means of the well known ordinary physical fields and, consequently, possesses energy of the interaction described by the interaction operator $\hat{V}(\vec{r}, t)$ (potential energy operator $\hat{V}(\vec{r})$, for conservative systems) in the Hamiltonian. Such a general understanding exhausts the widely accepted picture of the physical phenomena. If the quantum potential is considered as a real interaction, it should correspond to completely new physics that is not founded in the theory in advance. However, no any "unusual" physics, *e.g.*, interaction with unknown physical fields, the $\Psi$-field [5], is being considered for the quantum system and its relation with the physical reality at the formulation of the problem. Hence, the real existence of the quantum potential should be a subject to a lucky coincidence. Although the latter cannot be excluded, no direct evidences of existence of additional non-local interaction between given particle and the rest of the system (in fact, the rest of the universe) due to the quantum potential have been clearly shown up to now. The Bohm's idea of unified and undivided world is highly inspiring and beautiful but more convincing arguments are required so as to be accepted. That is why, the problem of the interpretation of the quantum mechanics will be considered within the usual understanding of the physical phenomena.

The appearance of the quantum potential is, in fact, a formal consequence from the particular way of description of the physical state accepted in the quantum mechanics, *i.e.*, by means of wave function of particular structure that obeys dynamical equation of particular structure. The quantum potential is responsible for all non-classical features in the quantum Hamilton-Jacobi equation. Without the quantum potential, it reduces to the classical Hamilton-Jacobi equation that determines the classical trajectory of the particle. An important indication of the meaning of the quantum potential consists in the fact that it is determined only by the amplitude $R$ of the wave function, $U(\vec{r}, t) = -(\hbar^2/2m)(\Delta R/R)$. The amplitude of the wave function has well recognized statistical meaning (probability density), not only in the Copenhagen interpretation, but also in the Bohmian mechanics [5]. Hence, the quantum potential introduces statistical properties in the quantum Hamilton-Jacobi equation and it can be referred to the probabilistic meaning of the amplitude of the wave function rather than to a real interaction. That is why, instead of quantum potential, it must be more reasonably called *statistical term*, as will be used hereafter. The statistical term includes the mass of the particle in the denominator. This enforces the statistical properties toward the microscopic objects, in agreement with the well known behavior of the quantum phenomena. The main quantum mechanical equation in the hydrodynamic presentation, Eq. (12), is an extended classical Hamilton-Jacobi equation by means of the statistical term, thus associating the statistical properties to the quantum dynamics. In this way, the quantum mechanics becomes dynamical and statistical theory simultaneously. The dynamical and the statistical properties of any quantum system become inseparable within the quantum mechanical description. This, in particular, explains why within the quantum mechanical description the conjugated dynamical variables of an individual quantum object, e.g., the coordinates and the respective components of the linear momentum, are determined not exactly, in general, and can be considered as hidden parameters in some approaches [5]. From an experimental point of view, the statistical properties arise from the fact that the dynamics of an individual quantum object is usually studied within a system of a large number of identical quantum objects due to of their, typically, microscopic nature. The quantum mechanical predictions, which results from a statistically averaged quantum dynamics, are in an excellent agreement with such an experimental approach [50].

Summarizing the theoretical and experimental arguments presented here, we cannot entirely accept either the Copenhagen or the ontological interpretation [5] of the quantum mechanics. In fact, the wave function is a complex construct from the amplitude and the phase that have different physical meaning. The action/phase is ruled by the real dynamical equation (12), the quantum Hamilton-Jacobi equation, which, beside of pure dynamical terms, includes in addition the statistical term. That is why, the action/phase has a quantum-dynamical meaning, which, however, is influenced by the statistical properties coming from the statistical term. The amplitude of the wave function, which is ruled by the continuity equation for the probability density, Eq. (13), has a statistical meaning, which is influenced by the dynamical properties of the action/phase through the term including the velocity, $\vec{v} = m^{-1}\nabla S(\vec{r}, t)$. Also, we consider that the amplitude of the wave function is related with objectively existing probability that is only registered by the observer, but not simply its knowledge (epistemology) about the state of the quantum object [51]. Finally, the statistical properties cannot be entirely associated with the randomization effect on the quantum system coming from the environment − interaction with other particles and fields, including the zero-point vacuum fluctuations. In some cases, the physical state consists of regularly distributed in space maxima and minima of the probability density that resemble the interference effects of the classical waves. This reveals the existence of additional mechanism that determines the way of motion of the quantum system, originally considered as a particle. The mechanism of such wave-like feature of a particle-like object, referred to as a particle-wave duality, remains yet unclear. Despite of such uncertainties, the above considerations show that the dynamical and the statistical properties of any quantum system take place simultaneously within the quantum mechanical description of the physical reality. The dynamical and the statistical properties are entangled [52] in the wave function by means of its phase and amplitude, respectively. Such relation between the elements of the wave function is clearly expressed within the hydrodynamic representation of the quantum mechanics but it is not apparent within the Schrödinger's representation. Existing in a given representation is sufficient to prove its existence in general and can be considered as a fundamental feature of the wave function. Accepting the MP causality does not require acceptance of the ontological interpretation, considering the wave function as a real physical field [5]. We consider that the wave function as a whole does not have a definite physical meaning, although its elements, amplitude and phase, are causally related with the physical reality and physical meaning can be attributed to both of them. Such an understanding lies between the Copenhagen interpretation (according to which the amplitude, but not the phase, has a physical (epistemological) meaning), and the deBroglie-Bohm causal interpretation (according to which, the entire wave function, being a real physical object/field, has physical meaning). Our understandings can be put into a kind of a *dynamics-statistical interpretation* of the wave function. Due to the causal dependence on the physical processes and the initial conditions, Sec.3.1, the dynamical properties of the phase/action (subject to abovementioned statistical influence) are fully recognized in the dynamics-statistical interpretation, but not only as "a generator of trajectories" [53]. It is remarkable that the dynamical and the statistical nature of the state become so well combined in the wave function that it gives the best known description of the physical reality at a quantum level.

Two main approaches of extraction of the physical information from the wave function can be distinguished, *probabilistic* and *dynamical* ones. The probabilistic approach, which is based on the $R \leftrightarrow$ *physical reality* relationship, employs a bilinear product of wave functions, the matrix element, as in the standard quantum mechanics. The MP dependence within this approach is not easy to be seen due to its appearance in the matrix elements in a product of exponential phase factors. This may wash out the phase information, particularly for the diagonal matrix elements. The $S / \Phi \leftrightarrow$ *physical reality* relationship suggests a dynamical way of extraction of the physical information. The dynamical approach, applied in the Bohmian mechanics [5], is based on the action/phase, from which the velocity/linear momentum and the trajectory of the particle can be determined. The dynamical approach clearly reveals the dependence of the physical processes on the MP. Due to the mutual dependence between the amplitude and phase, both ways of extraction of the physical information are not independent but become *complimentary* each other. On the other hand, according to Bohr's complimentary principle, there are two kinds of (complementary) pictures in the description of the physical reality, particle-like and wave-like. Both pictures can be distinguished here based on the elements of the wave function. The particle-like picture is primary related to the action/phase $S / \Phi$, influenced by the amplitude $R$ due to their mutual relation in Eq. (12). For example, a particle-like feature like the quantum trajectory can be introduced by the integration of the velocity $\partial_t \vec{r} = \vec{v} = m^{-1} \nabla S(\vec{r}, t)$ [5]. The wave-like picture is primary related with the amplitude $R$, influenced by the action/phase $S / \Phi$ due to their mutual relation in Eq. (13). The square of the amplitude determines the probability density distribution, which reveals the wave-like picture at suitably designed experiments. Each of these pictures can be stronger or lesser expressed depending on the particular conditions (see Appendix A). For example, the wave-like picture is completely dominating for the case of stationary states in the sense that the latter are continuously distributed within the space occupied by the quantum system and no sign of particle trajectories can be distinguished from the respective probability density distribution. From the other side, the particle-like picture becomes enforced in the case of superposition of stationary states that forms wave-packet. The wave-packets can be created in a way so as to be closely "tightened" around given particle, *e.g.*, intra-atomic electron [16] or molecular nucleus [17-19], and, although its motion does not depict the trajectory in the classical sense, it can trace definite details of the motion of the respective particle. The particle-like and the wave-like features of the quantum objects are well proved experimentally. On the other hand, no undisputable explanation of the *mechanism* of the wave-like behavior of such particle-like physical objects is proposed till now. The particle-like and the wave-like pictures become entangled [52] in the wave function through the phase/action and the amplitude. Such a point of view can be considered as a further extension of the Bohr's complimentary principle, this time, based on the elements of the wave function. As can be seen, the truth (as we believe) does not recognize the border between the different formulations/interpretations of the quantum mechanics and reasonable concepts can be found in most of them.

**5. Conclusion**

Theoretical and experimental evidences of existence of a fundamental relation between the phase of the wave function and the physical reality are presented systematically for the first time. Existence of material phase causality is conclusively established on that ground. The phase/action and the amplitude of the wave function are mutually related by coupled dynamical and continuity equations. The phase of the wave function is ruled by the real dynamical equation, the quantum Hamilton-Jacobi equation, modified statistically by the amplitude through the statistical term. Hence, the material phase is primary related with the dynamics of the quantum system. The amplitude of the wave function is ruled by the continuity equation of the probability density, influenced dynamically by the action/phase. Hence, the amplitude of the wave function is primary related with the statistical properties of the quantum system. The amplitude of the wave function is related with the real physical probability instead with the observer's knowledge (epistemology) about the state of the physical object. The wave function, as a whole, does not have a definite physical meaning (in agreement with the standard interpretation of the quantum mechanics) while each of its elements, the amplitude and the phase/action, are causally related with the physical reality and physical meaning can be attributed to each of them. In view of above, the material phase appears to be *a missed parameter* within the standard interpretation. The wave function incorporates the particle-wave duality through its phase and amplitude. The latter can be considered as a further extension of the Bohr's complementarity principle.

A new dynamics-statistical interpretation of the quantum mechanics is introduced, according to which the quantum mechanics is dynamical and statistical theory simultaneously. The dynamical and the statistical properties of any quantum system take place simultaneously in the quantum mechanical description of the physical reality. We believe that the dynamics-statistical interpretation puts the quantum mechanics to a more advance level, i.e., from being a tool of statistical predictions to a theory dominated by the quantum dynamics. This is still not the Einstein's dream of complete dynamical theory. Whether such theory will be created is a question addressed to the future.

**References**

[1] Dirac P A M 1958 *The principles of Quantum Mechanics* (Oxford, Clarendon Press)
[2] von Neumann J 1955 *Mathematical Foundations of Quantum Mechanics* (Princeton University Press)
[3] Feynman R, Leighton R and Sands M 1965 *The Feynman Lectures on Physics, Vol.3, Quantum mechanics*, (Addison-Wesley, Reading, MA)
[4] Einstein A, Podolsky B and Rosen N 1935 *Phys. Rev.* **47** 777
[5] Bohm D 1952 *Phys. Rev.* **85** 166
[6] Feynman R 1948 *Rev. Mod. Phys.* **20** 367
[7] Griffiths R B 1984 *Journal of Statistical Physics* **36** 219
[8] Everett H 1957 *Reviews of Modern Physics* **29** 454
[9] Zurek W H. 2003 *Reviews of Modern Physics* **75** 715
[10] Born M 1926 *Z. Phys.* **37** 863
[11] Landau LD and  Lifshitz E M 1989 *Quantum Mechanics, Nonrelativistic Theory* (Nauka, Moscow, in Russian)
[12] Aspect A, Dalibard J and Roger G 1982 *Phys. Rev. Lett.* **49** 1804
[13] Zeilinger A 1999 *Rev. Mod. Phys.* **71** S288
[14] Chapman M S, Hammond T D, Lenef A, Schmiedmayer J, Rubenstein R A, Smith E and Pritchard D 1995 *Phys. Rev. Lett.* **75** 3783
[15] Durr S, Nonn T and Rempe G 1998 *Nature London* **395** 33
[16] Noel M W and Stround Jr C R 1995 *Phys. Rev. Lett.* **75** 1252
[17] Scherer N F, Ruggiero A J, Du M and Fleming G R 1990 *J. Chem. Phys.* **93** 856
[18] Scherer N F et al. 1991 *J. Chem. Phys.* **95** 1487
[19] Gerdy J J, Dantus M, Bowman R M and Zewail A H 1990 *Chem. Phys. Lett.* **171** 1
[20] Koprinkov I G 2000 *Phys. Lett. A* **272** 10
[21] Koprinkov I G 2001 *J. Phys. B: At. Mol. and Opt. Phys.* **34** 3679
[22] Koprinkov I G 2003 *Proc. of the XXIII International Conference on Photonic Electronic and Atomic Collisions, ICPEAC (Stockholm, Sweden 23 - 29 July)* Mo196.
[23] Berry MV 1984 *Proc. R. Soc. London A* **392** 45
[24] Kienberger R, Goulielmakis E, Uiberacker M, Baltuska A, Yakovlev V, Bammer F, Scrinzi A, Westerwalbesloh Th, Kleineberg U, Heinzmann U, Drescher M, and Krausz F 2004 *Nature* **427** 817
[25] Sansone G, Benedetti E, Calegari F, Vozzi C, Avaldi L, Flammini R, Poletto L, Villoresi P, Altucci C, Velotta R, Stagira S, De Silvestri S and Nisoli M 2006 *Science* **314** 443

[26] Schultze M, Goulielmakis E, Uiberacker M, Hofstetter M, Kim J, Kim D, Krausz F and Kleineberg U 2007 *New J. of Phys.* **9** 243

[27] Brabec T and Krausz F 2000 *Rev. Mod. Phys.* **72** 545

[28] Bohm D 1952 *Quantum Theory, Chapters 6 and 22* (Prentice-Hall Inc., New York)

[29] C-Tannoudji C, Diu B and Laloë F 1991 *Quantum mechanics* (J. Wiley & Sons, N.Y)

[30]. The quantum mechanics is a deterministic theory because the quantum state $|\Psi\rangle$ evolves deterministically obeying the respective dynamical/Schrödinger equation, $\hat{H}|\Psi\rangle = i\hbar\partial_t|\Psi\rangle$. The abstract quantum state $|\Psi\rangle$, however, does not have a definite physical meaning and does not correspond to some known classical quantity, although quantities that have physical meaning can be constructed from it. On the other hand, the quantities that have well understandable meaning from a macroscopic point of view, *i.e.*, position, momentum, etc., cannot be followed strictly in a deterministic way, in general, but can be predicted only probabilistically. Due to such a reason, the quantum mechanics can be considered as a partial deterministic theory.

[31] The quantity $\Omega_0 / 2\sqrt{\tilde{\Omega}_0}$ is not extracted from the brackets in Eq. (6) in order to keep the form of the wave function similar to that one in Eq. (5)

[32] Some discussion will be useful in order to clarify the terminology used here. The dressed states arise from the simplest (within the classification considered here) *bare states* of an isolated quantum system (atom or molecule, *e.g.*, the atomic hydrogen states, Born–Oppenheimer type molecular states, etc.) interacting with an electromagnetic field. If the field changes adiabatically, *adiabatic dressed states* arise from the bare states once neglecting completely the nonadiabatic factors within the conventional adiabatic condition, see Ref. 33 of this work. The introduction of generalized adiabatic condition [20] allows constructing dressed states of more general form, in which the nonadiabatic factors (up to given order) from the field (time derivatives of the field amplitude and phase) and the dumping are retained. For complete treatment of the material phase effects, the initial material phase was also subject to consideration. Such states were called Phase Sensitive Adiabatic States or PSASs [20] because a (generalized) adiabatic condition is still required for their derivation. In contrast to other known adiabatic states, however, the PSASs explicitly contain nonadiabatic factors. As the latter have crucial importance for many physical problems, hereafter, these states will be more correctly called *Phase Sensitive Nonadiabatic Dressed States*, or *PSNADS*.

[33] Hanna D C, Yuratich M A and Cotter D 1979 *Nonlinear Optics of Free Atoms and Molecules* (Springer-Verlag)

[34] Born M and Fock V A 1928 *Zeitschrift für Physik A Hadrons and Nuclei* **51** 165

[35] Koprinkov I G 1997 *Proceedings of the Int. Conference on Lasers '97* (STS Press 1998) ed J J Carroll, T. A. Goldman pp. 291.

[36] de Broglie L 1926 C. R. Acad. Sci. Paris **183** 447

[37] von Madelung E 1926 Z. Phys. **40** 322

[38] In the Bohmian mechanics, the square of the amplitude of the wave function also has meaning of probability density.

[39] Schmutzer E 1973 *Grundprinzipien der classischen mechanik und und der calssischen feldtheorie (kanonisher apparat)* (VEB Deutscher Verlag der Wissenschaften, Berlin)

[40] Salières P, Carré B, Le Déroff L, Grasbon F, Paulus G G, Walther H, Kopold R, Becker W, Milošević D B, Sanpera A and Lewenstein M 2001 *Science* **292** 902

[41] Bellini M, Lyngå C, Tozzi A, Gaarde M B, Hänsch T W, L'Huillier A and Wahlström C-G 1998 *Phys. Rev. Lett.* **81** 297

[42] Chang Z, Rundquist A, Wang H, Christov I, Kapteyn H C and Murname M M 1998 *Phys. Rev. A* **58** R30

[43] O Scully M, Englert B-G, Walther H 1991 *Nature* **351** 111

[44] Chapman M S, Ekstrom C R, Hammond T D, Rubenstein R A, Schmiedmayer J, Wehinger S and Pritchard D E 1995 *Phys. Rev. Lett.* **74** 4783

[45] Richle F, Witte A, Kisters Th and Helmeke J 1992 *Appl. Phys. B* **54** 333

[46] Remetter T, Johnsson P, Mauritsson J, Varjú K, Ni Y, Lépine F, Gustafsson E, Kling M, Khan J, López-Martens R, Schafer K J, Vrakking M J J and L'Huillier A 2006 *Natute Physics* **2** 323

[47] Sterr U, K. Sengstock, J. H. Müller, D. Bettermann, W. Ertmer, Appl. Phys. B 54 (1992) 341

[48] Susskind L and Glogower J 1964 Physics **1** 49

[49] Pegg D T and Barnett S M 1989 *Phys. Rev. A* **39** 1665

[50] The quantum mechanics is the most successful theory for computation of the physical phenomena at a quantum level and it is considered that no contradictions exist with the experimental results. While we definitely agree with the first statement, the second one will be accepted after some comment. The experimentally observed phenomena agree with the predictions of the quantum mechanics if the experiment is done with a large number of quantum objects "at once", or a large number of subsequent experiments with single quantum objects superimposing the outcome of all individual observations/measurements. For example,

in the famous two-slit experiment, a single particle reproduces only "a point" of the whole interference picture predicted by the quantum mechanics. The latter can predict only the probability of the particle to hit given point of the screen. To reproduce the whole interference picture, one requires to superimpose a large number of individual interference events of single particles sent through the slits [13]. The conventional quantum mechanics does not reveal, in fact, the dynamics of a single particle and its behavior can be predicted only statistically.

[51] We consider that the wave function, respectively its amplitude, describes objectively existing reality but not the observer's knowledge about it. The evidence of that can be found in an especially designed EPR type experiments with entangled photons whose state of polarization is measured by arbitrary switched polarizers that remain completely unknown for the observer until the experiment is completed and the results are processed and stored, *e.g.*, G. Weihs, T. Jennewein, C. Simon, H. Weinfurter, A. Zeilinger, Phys. Rev. Lett. 81 (1988) 5039. The well known correlation between the polarization states of the entangled photons is found even taking such special precautions against the influence of the observer on the process of acquisition and processing the data during the measurement. This means that the EPR type correlations reflect properties of the physical reality itself but not the observer's knowledge about it. The latter entails the understanding that the wave function is related with objectively existing physical but not epistemological properties.

[52] The term "entanglement" is associated here with the appearance of the amplitude and the phase/action in the coupled equations (12) and (13), as well as with the fact that the wave function incorporates both, the dynamical and the statistical characteristics by means of its phase and amplitude, respectively. It differs from the notion of entanglement between the wave functions of different quantum systems, as in the case of EPR correlations, Schrödinger's cat states, etc.

[53] L. E. Ballentine, Quantum Mechanics: a Modern Development, (World Scientific Publishing Co. Pte. Ltd, 1998) p.397.

[54] T. S. Rose, M. J. Rosker, A. H. Zewail, J. Chem. Phys. 88 (1988) 6672

[55] A. Zellinger, R. Gähler, W. Treimer and W. Hampe, Rev. Mod. Phys. 60 (1988) 1067

**Caption to the figures:**

Fig.1. Bare (zero external field) (a) and dressed (non-zero external field) states (b) and (c) of the quantum system and the respective transitions at ground (b) and excited (c) initial conditions. The bold arrows show the optical pumping, the hollow arrows show transitions due to the nonadiabatic factors, and the wavy arrows show the dumping. The black spots show the initial state (initial condition) from which the process of excitation and population of the states begins.

**Appendix A**

The results in the preceding sections follow from strict theoretical and experimental arguments. In this and the subsequent Appendices, we present some considerations that, for now, cannot be supported by strong evidences but might appear useful in the future treatment of the related problems. In this Appendix, we will consider (at an intuitive level) how the concept of particle moving along a trajectory [5], and the related features, could be reconciled with some basic concepts in the quantum mechanics.

The concept of trajectory is not accepted within the standard quantum mechanics due to the limitations on the simultaneous determination of the coordinates and the respective components of the linear momentum coming from the Heisenberg's uncertainty relations. The solution of the quantum mechanical problems show that the physical state is substantially distributed in space with no apparent indications of trajectories. The agreement of the quantum mechanical predictions with the experimental results seems to support such an understanding. To our opinion, the lack of indications of trajectories in most of the cases could be a subject to suitable interpretation rather then to their real absence. For example, the concept of trajectory leads to an excellent quantitative agreement with the exact quantum mechanical results in some high-field phenomena, and, in addition, gives a clear pictorial explanation of the underlying physical mechanism [40]. The concept of trajectory is intuitively and logically clear, has a macroscopic correspondence and seems to be one of the preconditions to set the question about the physical picture and, on that ground, the physical mechanism of the quantum phenomena. Apart from the qualitative and the quantitative consistency, the pictorial representation of the physical processes and the relevant physical mechanism should be the final goal of any physical description and represents the most significant difference between the physics and the mathematics. The idea of trajectory can be brought to a quantitative level if consider that a particle moves along the trajectory in a way that depicts the probability density distribution $|\psi|^2$ ( $\left|\sum_i \psi_i\right|^2$, for a superposition state) and the associated averaged characteristics of the state if observed for long enough time. The probability density distribution in a given space "point" can be related with the relative time spent by the particle in that "point". The electron orbiting time, say, for the ground state of the hydrogen atom is about 150 attoseconds [24]. Hence, even some typical microscopic time scales, *e.g.*, the lifetime of an excited electronic state, which is in the order of nanoseconds or much longer, may become long enough to enable the electron motion along the trajectory to depict rather well the probability distribution. Still more, this holds for the stationary states, the most widely considered states in the quantum mechanics, that have infinitely long lifetime. In the same line of reasoning, the "instantaneous" collapse of the wave function during the measurement can be considered as a fast, but continuous in time, rearrangement of the way of motion (and the associated averaged characteristics) corresponding to the superposition state into a way of motion (and the associated averaged characteristics) corresponding to the measured state.

The space distributed states result from the solution of the quantum mechanical problems in terms of the wave function that is distributed in space. The main challenge is to explain the space distributed state in reality based on the concept of trajectory. A partial explanation is that the space-distributed states result from perturbations coming from the environment. The universal nature of the quantum perturbations becomes clear, for example, from the limited lifetime of any stationary excited state of even "isolated" atom/molecule (that should be infinitely long, as for the stationary states of the Hermitian Hamiltonians) due to, at least, the zero-point vacuum fluctuations as an universal perturbation. The existence of substantial perturbations is an inherent feature of the quantum phenomena due to the small mass of the typical quantum objects. The typical characteristics of the quantum systems, *e.g.*, the microscopic size, impose, in most of the cases, performing the experimental studies with a large number of identical quantum systems simultaneously. In more refined, in general, experiments, superposition of a large number of experimental results with single quantum systems is also applied. To our opinion, the quantum mechanical results (*e.g.*, the space distributed states) can be understood in terms of trajectories if consider a superposition of all possible real trajectories associated with the external motion of the quantum systems, for the case of external states, and/or the internal motion of the building parts of the quantum system, for the case of internal states, taking into account the perturbations. It has randomization effect on the electron motion and may completely wash out the effect of the individual trajectories. The states formed by strongly randomized trajectories entail statistical description, as in the usual quantum mechanics. On the other hand, a strong external field of given polarization (comparable or, sometimes, even stronger than the intra-atomic field) may peak up from the background of the randomized trajectories a relatively low number of electron trajectories of strongly dominating contribution [40]. It enforces the particle-like picture and entails dynamical description. The action of the randomization factors, however, cannot explain the regularly distributed maxima and minima of the probability distribution, and, particularly, the existence of selected types of quantum states. Although the latter have their formal solution within the Schrödinger equation, the physical mechanism of creation of regularly distributed selected type of states remains unclear. It is also unclear how to reconcile the local concept of particle moving along a trajectory with the non-local phenomena as the interference of single particle (Feynman's "which-way" interference experiment) and the EPR type correlations. In any case, recent advances in the attosecond metrology [24-26] set the question of the real time dynamics of atomic/molecular electrons, as it was done some time ago with the nuclear motion inside the molecules [54].

The particle-like features (*e.g.*, space localization, trajectory, etc.), to our opinion, should be accepted as a primary fundamental nature of a single physical object, while the wave-like features (*e.g.*, space distributed state, non-local behavior, interference, etc.) should be considered as a simultaneous or consecutive collective effect of many individual quantum systems. Such an understanding can be supported by few arguments: (*i*) a single particle is always localized in the experiments of its detection, (*ii*) no wave-like phenomenon with a single particle in a single experiment has been observed but only in a simultaneous experiment with a large number of identical particles, or in a large number of consecutive experiments with single particles. Really, the experiments with low-intensity fields show that the interference picture is formed "point" by "point" after each particle (quantum of the field, or more complex object - atom, molecule, etc.) hits the detector, e.g., phosphorescent screen [55]. At that stage, the particle manifests itself as a localized object, nevertheless that its actual dimension is subject to additional investigations. Once the interference picture is completed, it reveals the wave-like feature of the same physical object, more correctly, a large number of identical physical objects that form together the interference picture. Hence, the interference phenomena are capable to reveal not only the wave-like features, but also some aspects of the particle-like features. The second argument above can be reconciled with the Dirac's understanding that a single particle interfere by itself [1] if consider that the mechanism responsible for the interference of a single particle acts in any such event involving single particle, whereas the interference picture is proved after a superposition of a large number of such events. The main problem with the particle-wave duality is to reconcile in a single physical object so opposite features like the particle-like and wave-like ones, unless we assume that the particle is involved in a wavelike motion due to the action of some universal physical background.

It is important to emphasize that the MP causality established here is not related with the validity of the concept of trajectory and the related properties. The latter would only help to build a more clear pictorial presentation of the quantum phenomena. To what extend, if ever, the concept of trajectory can be forwarded to encompass entirely the quantum phenomena is a question addressed to the future.

**Appendix B**

The electromagnetic waves can well be described in terms of classical wave, at least for high-intensity fields. In some aspects, however, they are closer to the matter (de Broglie) waves and they both strongly differ from the typical classical waves as, *e.g.*, elastic/sound waves. The elastic waves represent collective oscillations of a large number of "classical" particles of variable phase delay, joined by elastic forces. Such waves cannot be formed if the number (number density) of the involved particles is too low, and, consequently, the elastic forces between the particles are negligible. This kind of waves will be generally called *oscillating waves* and will be considered as classical waves. In contrast to the oscillating waves, the electromagnetic and the matter waves retain their wavelike features at very low density of the propagating particles, including even the case of single particles. That is why, these waves do not represent collective oscillations of particles but they arise from the specific way of propagation of the particles/quanta of the respective fields (photons, electrons, neutrons, atoms, etc.). Such propagating particles are considered as waves in some phenomena *e.g.*, interference and diffraction, because they manifest a behavior that is similar to that of the usual oscillating waves and, from a given point of view, can be explained in terms of waves. Such kind of waves will be generally called *particle waves*.

The wavelike behavior is usually associated with a kind of oscillatory field. The implementation of cutting edge ultrafast laser technology, namely, extreme ultraviolet single attosecond pulses and infrared high-intensity femtosecond laser pulses, in some resent experiments [24] seem to confirm (at least, for the high intensity/number density fields) the oscillation-like behavior of the optical field. It was shown that the released electrons due to the ionization of atoms by an attosecond pulse obtain oscillation-like momentum distribution under the action of another pulse of femtosecond duration. The oscillation field (apart from the fact that the solution of the respective wave equations is oscillation-like) seems to be the most natural explanation of the origin of an oscillation force on a charged particle. To our opinion, the oscillation-like momentum distribution of the released electrons is not enough to make the ultimate conclusion for the oscillation-like nature of the optical field (although it might happen that it is really the case), but it is rather an indication that the optical field exerts an oscillation-like force on the electrons that results in an oscillation-like momentum distribution. The origin of an oscillation force on charged particle in an electromagnetic field considered as a flux of quanta/particles seems not trivial, especially at low-density or single-quanta fields.

**Appendix C.**

The phase/action is not simply one of the physical observables, but one may find a more distinguished role of this quantity in physics, including both, the particle-like and the wave-like pictures of the physical reality. In the wave-like phenomena, the phase/action of the state of a physical object, *e.g.*, vectors (electric and magnetic) of the electromagnetic field, the wave function of a particle, etc., plays a crucial role, together with the amplitude, of course. In the particle-like phenomena, the main role is performed by the dynamical quantities that arise from the phase/action by derivatives: velocity $\vec{v} = m^{-1}\nabla S$, linear momentum $\vec{p} = \nabla S$, Lagrangian $L = \partial_t S$, Hamiltonian (energy, for conservative systems) $H = -\partial_t S$ ($E = -\partial_t S$), etc. Being an integral quantity, the phase/action does not give direct information about the local/instantaneous behavior of the quantum system, as, *e.g.*, the linear momentum, energy, etc. do, although these quantities can be obtained considering the rate of change of the phase/action. On the other hand, the "local" quantities lose the relevant physical information carried by constant values of phase/action, or functions not depending on the variable of the respective derivative.

The phase (action), $\Phi(t) = \hbar^{-1}\int_0^t H dt'$ ($S(t) = -\int_0^t H dt'$), is an integral quantity, that acquires contributions from all events (motions and interactions) that have happened to the quantum system. This constitutes the *physical history* of the quantum system. Any motion (does not matter inertial or accelerated) and interaction of the quantum system results in a definite contribution to the phase/action, additively to its initial value. It invokes the understanding that the particle moves so that the MP is "rolling without sliding on the space". Any such contribution can be made observable if it is compared interferometrically with other reference state. Thus, the simple motion of a particle across the empty space seems not to be a trivial matter. Even for an isolated quantum system at rest, the phase (action) makes "idle" running in time thus acquiring a definite value, $\Phi_0(t) = \hbar^{-1}\int_0^t H_0 dt'$ ($S_0(t) = -\int_0^t H_0 dt'$). Here, $H_0$ is the Hamiltonian of free quantum system at rest, which is non-zero due to the internal motions and interactions of the building parts of the quantum system. In principle, such "idle" running of the phase/action could be made observable at suitably designed interferometric experiment. The time rate of the phase/action acquisition, *i.e.*, frequency/energy, is ruled by $H_0$. Such a rate can be considered as a kind of *internal time/"clock"* of the quantum system.